\documentclass[a4paper]{article}

\usepackage{icrc2013}

\title{Unprecedented temporal evolution of the broad-band emission of the BL Lac Mrk~501 }

\shorttitle{MWL variability of the BL Lac Mrk~501}

\authors{
Marlene Doert$^{1,3}$,
David Paneque$^{2}$,
for the MAGIC Collaboration, the VERITAS Collaboration and the \textit{Fermi}-LAT Collaboration.
}

\afiliations{
$^1$ Technische Universit\"{a}t Dortmund, 44221 Dortmund, Germany.\\
$^2$ Max Planck Institut f\"{u}r Physik, 80805 M\"{u}nchen, Germany.\\
\scriptsize{
$^3$ now at: Columbia University, New York, NY 10027, USA.\\
}
}

\email{mdoert@nevis.columbia.edu}

\abstract{The TeV BL Lac object Markarian 501 has been the subject of a 4.5 month multi-instrument campaign conducted in 2009, which provided an excellent temporal and energy coverage from radio to very high energy gamma rays ($>$100\,GeV, VHE). During the campaign, Mrk~501 was mostly in a comparably low state, but for two flares at VHE with very different characteristics. While the second flare seems to be correlated with a moderate increase in the X-ray flux, the first one is most likely accompanied by a shift of the synchrotron bump towards higher energies. Moreover, the first flare occurs during an abrupt change in the polarized optical flux, and was preceded by a rotation of the electric field vector position angle. This is the first time that such behavior is observed in a high-frequency-peaked BL Lac object, while similar events have been seen in the low-frequency peaked BL Lac object BL Lacertae and the flat spectrum radio quasar PKS~1510-089, hence suggesting that similar physical processes occur in the jets of different blazar subclasses.}


\begin{document}
\maketitle

\section{Introduction}
Due to its proximity ($z=0.034$) and its frequent episodes of high activity, Mrk~501 is considered among the best sources to study the intrinsic mechanisms which cause the highly variable broad-band emission of blazars. Here we present results from a multi-instrument campaign on this object, which allowed to study the average spectral energy distribution (SED) of the source's output as well as temporal changes in the emission. While the newly gained insights on the average state emission have already been published in \cite{Abdo2011}, the source's variability is addressed here and will be presented in detail in a forthcoming paper.

\section{Multi-instrument campaign}
The presented multi-wavelength (MWL) campaign was conducted over 4.5 months in 2009. The aim of this campaign was to sample the SED over all wavelengths every $\approx5$ days. This way, the intrinsic flux variability of the source could be probed without any bias towards high flux states. Data were taken by 30 different instruments, from Radio to VHE gamma rays, between 2009 March 15 (MJD~54905) and 2009 August 1 (MJD~55044). Good temporal coverage was achieved, while the sampling density naturally varies between the different wavebands. 
For details on the observation strategy, list of instruments and analysis procedures performed for the different instruments, the reader is referred to \cite{Abdo2011} and references therein.\\
The initial MWL data set was expanded by optical polarization measurements performed by the Steward Observatory in April and May 2009.

\section{MWL flux variability}
\begin{figure*}[t]
  \centering
\includegraphics[width=0.85\textwidth]{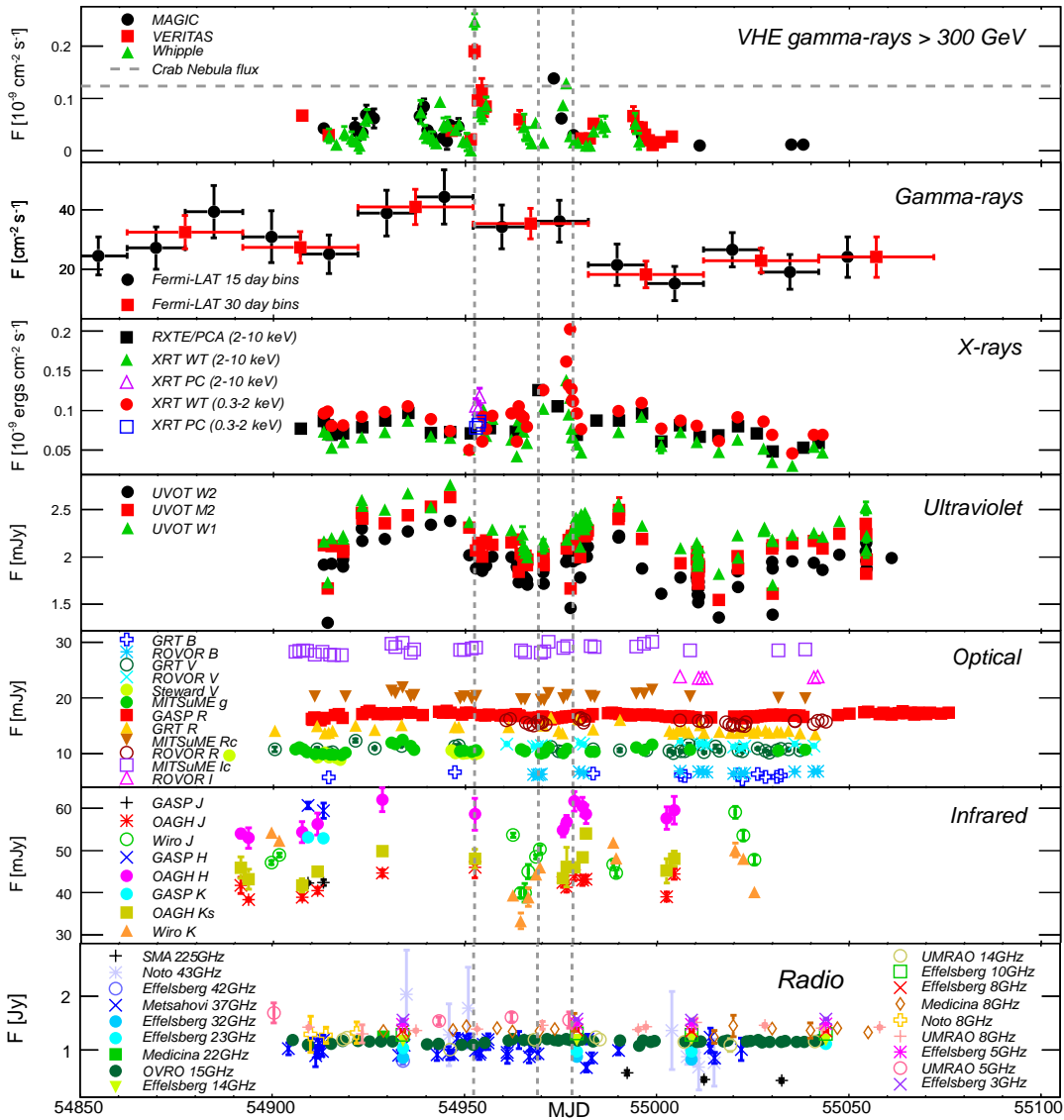}
  \caption{Light curves in the various energy bands. Whipple 10-m fluxes were scaled to report a flux $>$300GeV, while the actual energy threshold is 400 GeV. Vertical lines indicate a sharp VHE flare and a period of increased VHE activity.}
  \label{fig:mwllcs}
 \end{figure*}

\begin{figure*}[t]
  \centering
\includegraphics[width=0.8\textwidth]{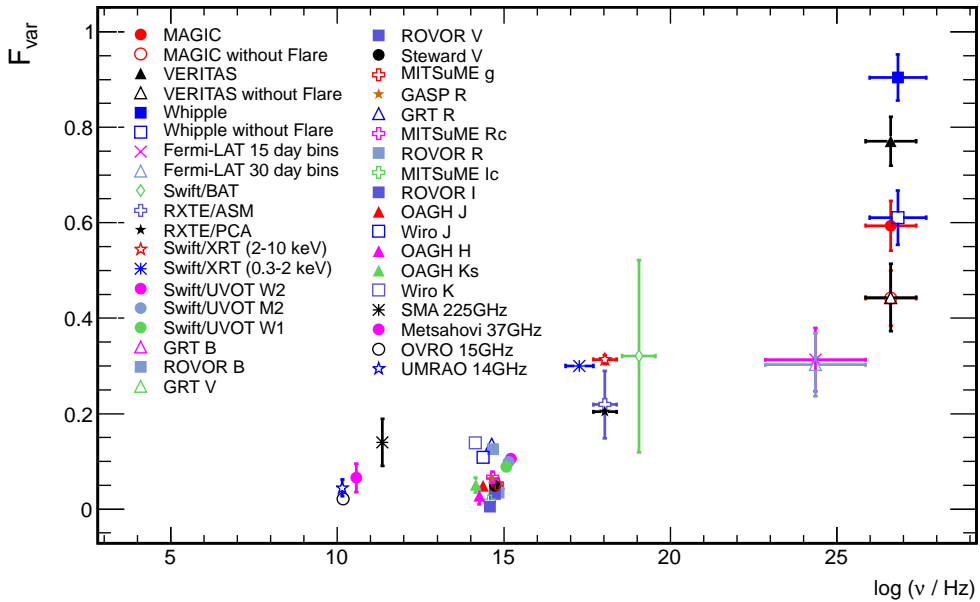}
  \caption{Fractional variability for light curve measurements at different frequencies, following \cite{Vaughan:2003iu}.}
  \label{Fig:fvar}
 \end{figure*}

The light curves from all instruments which deliver single-day flux points are shown in Fig.~\ref{fig:mwllcs}.\\
In the VHE gamma-ray band, Mrk~501 was mostly in a comparably low state at a level of $\approx 0.3$ times the flux of the Crab Nebula, which we denote by Crab Units (C.U.; 1 C.U. corresponds to a flux $>300$ of 1.2 x $10^{-10}$ cm$^{-2}$ s$^{-1}$) during the 4.5 month long campaign. Despite the overall low flux, variability was seen. Two VHE flares were covered during this campaign. One occurred on MJD~54952 and another around MJD~54973. \\
In the \textit{Fermi}-LAT measurements, significant variation can be seen, where the largest flux is observed in the time interval from MJD~54952 to MJD~54982. Short-term variability could not be probed based on these measurements due to integration times of 15 and 30 days for the light curve.\\ 
While \textit{RXTE}'s PCA and ASM measured only small flux variations in the X-ray regime, significant variability was seen with \textit{Swift}/XRT, both below and above 2~keV, as well as with BAT. The differences in variability between the X-ray instruments is caused by the different temporal coverage. Besides variations of 50 to 60\% in flux on time scales of 10 to 20 days, \textit{Swift}/XRT also covered a large flux increase in both energy bands around MJD~54977. \\
The GASP R-band observations provide a very well sampled light curve, which shows small variations of about 10\% in flux over time scales of about 15 days.\\
In the ultraviolet, significant flux variations of about 25\% over time scales of about 25 to 40 days can be seen.\\
Only small variations can be seen in the near-infrared measurements. For the extensive data sample in the optical regime, a nearly constant flux was also measured, with only small variations, e.g.~around MJD~54935 and MJD~55000, apparent in the Mitsume data (in all bands). 
A nearly constant flux can be noted in the radio regime. \\
As a quantitative study of the underlying variability seen at different wavelengths, the fractional variability $F_{var}$ has been determined for each instrument, following \cite{Vaughan:2003iu}. The results are presented in Fig.~\ref{Fig:fvar}.
At low frequencies, from radio to optical, no strong variability was detected, with an $F_{var}$ of about $0.05$ in radio and $\approx 0.20$ in the optical regime. In the X-ray band, $F_{var}$ values of the order 0.3 are obtained, indicating some variation in the flux during the probed time interval. The fractional variability in the high energy gamma-ray band covered by \textit{Fermi}-LAT is of the same order. It has to be noted that due to the larger integration times of the \textit{Swift}-BAT (30 days), \textit{RXTE}/ASM (30 days) and \textit{Fermi}-LAT (15/30 days) measurements, the $F_{var}$ values are not directly comparable to the other instruments, as short time variability can not be probed. In VHE gamma rays, we show the fractional variability for both the original light curves and light curves for which the flare(s) have been excluded. For all light curves, strong variability can be noted, with $F_{var}\geq 0.4$ for the non-flare light curves and $F_{var}\geq 0.6$ (0.9 for the Whipple 10-m) for observations including the flaring episodes.\\
All in all, Mrk~501 showed a large increase in variability with increasing energy, ranging from a steady behavior at the lowest frequencies to strong variations in flux at the highest energies. This is in contrast to the behavior seen from Mrk~421 in 2009, where the highest variability occurs in the X-ray band, as reported in \cite{2012AIPC.1505..518N}.

\begin{figure}[t]
  \centering
\includegraphics[width=0.5\textwidth]{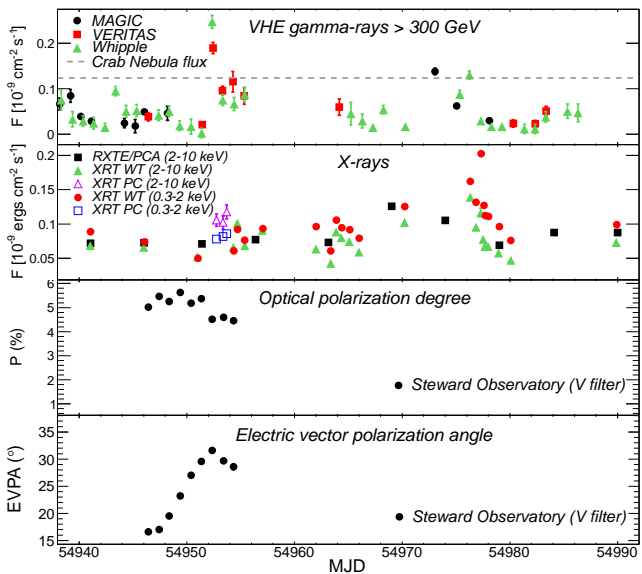}
  \caption{Zoomed in light curve around the periods of VHE flaring. Whipple 10-m fluxes were scaled to report a flux $>$300GeV, while the actual energy threshold is 400 GeV. }
  \label{fig:closeupLC}
\end{figure}

\noindent
The two flares which occurred in the course of the presented campaign have been analyzed in detail regarding flux variations and spectral changes. 
Fig.~\ref{fig:closeupLC} shows a close-up view around the time of the flaring events of the light curves obtained in VHE gamma rays by MAGIC, VERITAS and the Whipple 10-m, in X-rays by \textit{Swift}/XRT and \textit{RXTE}/PCA, and the degree of optical polarization and the corresponding electric field vector position angle (EVPA) as measured by the Steward Observatory.

\noindent
On MJD~54952, the Whipple 10-m measured a flux increase up to a peak flux of 4.5\,C.U. and an average flux of 2\,C.U., which is about one order of magnitude higher than the low flux level which was observed during this campaign, over an observation time of 2.3 hours. VERITAS observed simultaneously with the Whipple 10-m for the last 1.4 hours and recorded an average flux of 1.5\,C.U.~(about 6 times the low flux level). The rise time in which the flux increased by a factor 5 was measured by the Whipple 10-m as 25 minutes, indicating fast variability \cite{2011ICRC....8..171P}.\\
Dedicated spectra have been derived based on the high state observations, which show a tentative hardening compared to the ``low state'' spectra \cite{2011ICRC....8..171P}.\\
At the time of the VHE flare, no significant increase of the X-ray flux can be detected in the \textit{Swift}/XRT observations. However, it has to be noted that the observations are not strictly simultaneous to the VHE data and started 7 hours after the Whipple 10-m and VERITAS observations. No \textit{RXTE}/PCA measurements took place around that time.\\
The X-ray spectra obtained with the XRT show a significant hardening around the VHE outburst with an upward curvature of the spectral distribution. The spectral index as obtained from a simple power-law fit is found to be $-1.75$ during the time of the flare, in contrast to the average spectral index during the campaign of $-2.05$. This hardening indicates a likely shift of the synchrotron peak towards higher energies.\\
While the degree of optical polarization is comparatively high shortly before the VHE flare, it drops by $\approx 15 \%$ at the time when the flare occurs. The EVPA shows a rotation by 15 degrees, which comes to a halt at the time of the VHE outburst. Such a coincidence of a gamma-ray flare and a break in the evolution of the optical polarization has never been seen before for a high-frequency-peaked BL Lac (HBL) object. \\
\\
On  MJD~54973, MAGIC saw an increased VHE flux of 1.1 C.U., which corresponds to nearly 4 times the low flux level. The data were probed for variations on time-scales down to minutes, but no significant intra-night variability was found. The Whipple 10-m saw increased fluxes two and three days later (MJD~54975 and MJD~54976). A ``high state'' spectrum was derived from the MAGIC data, which also shows a tentatively harder slope.\\
Also for this event, no strictly simultaneous X-ray observations are at hand. The closest \textit{RXTE}/PCA observations took place 4 days before and one day after the MAGIC exposure. On both days, only a slight flux increase was seen. \textit{Swift}/XRT observed on MJD~54976.3, quasi-simultaneous to the flux increase in VHE seen by the Whipple 10-m, and saw an enhanced X-ray flux in both energy bands. In the subsequent, short X-ray flare seen by \textit{Swift}/XRT on MJD~54977, the flux increased to more than 2 times the typical flux level during this campaign. The rise time of the flare cannot be stated properly due to a gap in the XRT measurements between MJD~54970 and MJD~54976, but the decay time of the flare is on the order of days. The X-ray spectra show some variation, but unlike the first event no particular hardening is visible.\\
Despite the non-simultaneity of the observations in the different bands, it can be noted that the source was in a state of increased activity over a period of up to 7 days.

\section{Discussion}
A trend of increasing variability for increasing energies within the X-ray and VHE band in prior observations of Mrk~501 have been reported previously \cite{Gliozzi:2006it,Albert:2007bt}. In the work presented here, we show that this trend occurs throughout all the energy bands, from radio to VHE, culminating in fast and strong variability at the highest energies. The occurrence of high energy flares from Mrk~501 is a well-known phenomenon. Still the flaring events seen here allow for some new conclusions: The first flare is dominated by a fast outburst in the VHE band, which has not been accompanied by a significant increase of the X-ray flux. Based on these observations, this event was tentatively categorized earlier as an ``orphan flare'' \cite{2011ICRC....8..171P,2012A&A...541A..31N}, which is a conclusion that would substantially challenge the SSC emission models currently favored for HBL. Based on the observed SED at the time of the flare, the data more likely indicate a shift of the synchrotron bump towards higher energies, which is suggested by the steep increase of the XRT-spectrum. During the outstanding flare in 1997, the synchrotron bump has been observed to be shifted to beyond 100\,keV \cite{1998ApJ...492L..17P}.
\\
The observed change in the optical polarization in coincidence with the TeV flare 
generally suggests a scenario of particle injection into the emission region or the encounter of a turbulent plasma region in the jet, where the (partly ordered) movement of the emitting particles in the surrounding magnetic field is disturbed.\\
The second flare, which occurred around MJD~54973, shows a different appearance in terms of flux levels, the shape of the SED and the involved time scales. Contrary to the behavior seen around MJD~54952, Mrk~501 showed a level of increased activity and erratic flux changes over several days. This could be explained by the passage of the emission zone through an extended shock region, which would cause an increase of the particle density through the compression of the region itself.\\
While the polarization behavior discussed during the flaring events has never been seen before for an HBL, similar observations have been made for the case of the low-frequency peaked BL Lac (LBL) object BL~Lacertae and the flat spectrum radio quasar (FSRQ) PKS~1510-089, where a change in the optical polarization also emerged in coincidence with the first of a pair of flares (or with the start of a sequence of flares) \cite{Marscher:2008ii,Marscher:2010hu}. For the case of BL~Lac it was suggested that the first event occurs at the time when the emission region travels along the last spiral arm of a helical path within the acceleration and collimation zone of the jet and finally leaves this zone to enter a more turbulent region, while the second flare is interpreted as the passage of the region through a standing shock located at the radio core \cite{Marscher:2008ii}. \\
Based on the light curves, Mrk~501 shows several similarities to the discussed behavior in PKS~1510-089 and especially in BL~Lacertae. Exhibiting different peak frequencies for the synchrotron and the IC bump, the optical flaring activity observed in BL~Lac can be seen as corresponding to the X-ray variability in Mrk~501. The observed degree in optical polarization in Mrk~501 $(\approx 5\%)$ appears to be small in comparison to BL~Lacertae (up to $18\%$). However, the optical flux in Mrk~501 is strongly dominated by the host galaxy, so that the jet contribution amounts to only 25-30\%. Therefore, the measured degree of polarized light corresponds to  a fraction of $\approx 15-20\%$ of polarized emission from the jet, which is comparable to BL~Lacertae. The striking similarities in the observed behavior of the different sources suggest that the scenario discussed in \cite{Marscher:2008ii} could also be applicable to the present observations of Mrk~501. Furthermore, they give further evidence to the idea of a intrinsic similarity between the different sub classes of blazars, despite their obvious differences such as the shape of the SED and the total jet power.\\
Further details and a discussion on temporal changes in the broad-band SED of Mrk~501 will be presented in a forthcoming publication.

\vspace*{0.5cm}
\footnotesize{{\bf Acknowledgment:}{\\
The MAGIC Collaboration would like to thank the Instituto de Astrof\'{\i}sica de
Canarias for the excellent working conditions at the
Observatorio del Roque de los Muchachos in La Palma.
The support of the German BMBF and MPG, the Italian INFN, 
the Swiss National Fund SNF, and the Spanish MICINN is 
gratefully acknowledged. This work was also supported by the CPAN CSD2007-00042 and MultiDark
CSD2009-00064 projects of the Spanish Consolider-Ingenio 2010
programme, by grant 127740 of 
the Academy of Finland,
by the DFG Cluster of Excellence ``Origin and Structure of the 
Universe'', by the DFG Collaborative Research Centers SFB823/C4 and SFB876/C3,
and by the Polish MNiSzW grant 745/N-HESS-MAGIC/2010/0.\\
This research is supported by grants from the U.S.
Department of Energy Office of Science, the U.S. National
Science Foundation and the Smithsonian Institution, by
NSERC in Canada, by Science Foundation Ireland (SFI
10/RFP/AST2748) and by STFC in the U.K. We acknowledge
the excellent work of the technical support staff at the Fred
Lawrence Whipple Observatory and at the collaborating institutions
in the construction and operation of the instrument.\\
The $Fermi$ LAT Collaboration acknowledges support from a number of agencies and institutes for both development and the operation of the LAT as well as scientific data analysis. These include NASA and DOE in the United States, CEA/Irfu and IN2P3/CNRS in France, ASI and INFN in Italy, MEXT, KEK, and JAXA in Japan, and the K.~A.~Wallenberg Foundation, the Swedish Research Council and the National Space Board in Sweden. Additional support from INAF in Italy and CNES in France for science analysis during the operations phase is also gratefully acknowledged.}}

\end{document}